\documentclass[twocolumn, astrosymb, usenames, dvipsnames]{aastex631} %

\newcommand{\identityone}[1]{#1}
\newcommand{\identitytwo}[1]{#1}

\usepackage{color}
\usepackage{xcolor}
\usepackage{amsmath} 
\usepackage{gensymb} 

\newcommand{\gx}{GX~339$-$4}

\newcommand{\beq}{\begin{equation}}
\newcommand{\eeq}{\end{equation}}

\newcommand{\approptoinn}[2]{\mathrel{\vcenter{
  \offinterlineskip\halign{\hfil$##$\cr
    #1\propto\cr\noalign{\kern2pt}#1\sim\cr\noalign{\kern-2pt}}}}}

\newcommand{\appropto}{\mathpalette\approptoinn\relax}

\shortauthors{Pjanka et al.}
\graphicspath{{./}{figures/}}

\begin{document}

\title{Shock corrugation to the rescue of the internal shock model in microquasars\\
The single-scale MHD view}

\correspondingauthor{Patryk Pjanka}
\email{patryk.pjanka@gmail.com}

\author[0000-0003-3564-9689]{Patryk Pjanka}
\affiliation{Nordita, KTH Royal Institute of Technology and Stockholm University, Hannes Alfvéns väg 12, SE-106 91 Stockholm, Sweden}

\author[0000-0003-3433-0772]{Camilia Demidem}
\affiliation{\identityone{JILA, University of Colorado and National Institute of Standards and Technology,\\ 440 UCB, Boulder, CO 80309-0440, USA}}
\affiliation{Nordita, KTH Royal Institute of Technology and Stockholm University, Hannes Alfvéns väg 12, SE-106 91 Stockholm, Sweden}

\author[0000-0002-5767-7253]{Alexandra Veledina}
\affiliation{Department of Physics and Astronomy, FI-20014 University of Turku, Finland}
\affiliation{Nordita, KTH Royal Institute of Technology and Stockholm University, Hannes Alfvéns väg 12, SE-106 91 Stockholm, Sweden}


\begin{abstract}

Questions regarding energy dissipation in astrophysical jets are open to date, despite of numerous attempts to limit the diversity of models.
Some of the most popular models assume that energy is transferred to particles via internal shocks, which develop as a consequence of non-uniform velocity of the jet matter.
In this context, we study the structure and energy deposition of \identityone{colliding plasma shells}, focusing our attention on the case of initially inhomogeneous shells.
This leads to formation of distorted (corrugated) shock fronts -- a setup that has recently been shown to revive particle acceleration in \identityone{relativistic magnetized perpendicular shocks}.
Our studies show that the radiative power of the far downstream of \identityone{non-relativistic magnetized perpendicular} shocks is moderately enhanced with respect to the flat shock cases.
Based on the decay rate of downstream magnetic field, we make predictions for multiwavelength polarization properties.

\end{abstract}


\section{Introduction}\label{sect:intro}

Fast relativistic outflows -- jets -- are common to astrophysical environments harboring accretion and~/~or explosions.
Sources standing orders of magnitude apart in energy dissipation, such as young stellar objects, accreting black hole X-ray binaries, gamma-ray bursts, active galaxies and blazars, have all been witnessed to launch such outflows \citep[e.g.,][]{1997Goodson, MirabelRodriguez1999, 1999MacFadyen}.
The mechanisms behind jet acceleration and fueling are, however, debated \citep{Komissarov2011,DavisTchekhovskoy2020}.
Jets are known to produce bright non-thermal emission, often observed as power-law synchrotron and/or inverse Compton continua \citep[e.g., review by][]{Matthews2020}, indicating continuously-operating energy dissipation and particle acceleration mechanisms.

Jet kinetic energy can be converted to radiation via magnetic reconnection events \citep{Lyutikov2003}, pair processes \citep{Derishev2003,SternPoutanen2008} or shocks in the interior of the jets \citep[]{Peer2014}.
In the latter scenario, the jet is thought to be ejected as inhomogeneous matter, both in density and in velocity. This gives rise to discrete pockets of plasma (shells) travelling through the jet at different speeds. Once a faster shell catches up with a (previously ejected) slower one, they collide, and two shocks are formed (under the right conditions, see \citealt{Peer2014}): a forward shock driven into the slower shell, and a reverse shock driven into the faster one. These shocks could enable particle acceleration and, consequently, power the non-thermal emission of the jet. After the collision, the shells are typically assumed to merge inelastically and continue propagating as a single shell (e.g., \citealt{Beloborodov2000, 2014Malzac}, although see \citealt{Kino2004}). They then encounter other shells and collide again, ensuring continuous particle re-acceleration throughout the jet. 
This ``internal shock model'' has been originally proposed in the context of the resolved inhomogeneous structures, the so-called knots, in the jet of M87 galaxy \citep{Rees1978}.
Later, the model was generalised to jets observed from other astrophysical sources: radio-loud quasars \citep{Spada2001}, gamma-ray bursts \citep{ReesMeszaros1994,MedvedevLoeb1999,Beloborodov2000}, and microquasars \citep{Kaiser2000,2014Malzac}.

Such bright knots have been observed in a number of sources, supporting localised energy dissipation scenarios \citep{Matthews2020}.
The observed collision of these knots in the radio galaxy 3C~264 \citep{Meyer2015} gave observational confirmation to the internal shock scenario in context of AGN jets.
In microquasars, internal shock models have proved to be successful in explaining both spectral and timing properties \citep{Vincentelli2019,2019Peault}.
However, such modelling relies on a number of assumptions about energy liberation and particle acceleration in shell collisions, which need to be verified.

Different aspects of the flow structure and jet observables in the presence of internal shocks have been considered: the geometrical and hydrodynamic evolution \citep{Kino2004, Mimica2005, 2012Granot, 2020Marino, 2020Rudolph}, thermodynamics \citep[][]{2001KobayashiSari, Mimica2004, 2008Graff, 2017Peer}, the impact of magnetic fields \citep[][]{2004FanWeiZhang, 2007Mimica, 2010MimicaAloy, 2012Mimica, 2014RuedaBecerril, 2015RuedaBecerril, 2015Deng}, radiative processes \citep[][]{2008Graff, BoettcherDermer2010, 2011JoshiBottcher, 2012JamilBottcher}, and timing properties \citep[][]{Spada2001, Mimica2005, BoettcherDermer2010, 2010Jamil, 2013Malzac, 2014Malzac, 2015Drappeau}, among others. While the internal shock scenario is quite successful in representing a number of sources \citep[e.g.,][]{1998Daigne, Spada2001, 2015Drappeau, 2018_Malzac, 2019Peault, 2020Bassi, 2020Marino}, its main problem (and its main criticism) lies in its apparent inefficiency of energy conversion from kinetic to that of non-thermal particles (though this may depend on model assumptions, as pointed out by \citealt{2017Peer}).
It has been shown \citep{BegelmanKirk1990,SironiSpitkovsky2009,SironiSpitkovsky2011} that electrons can consistently achieve non-thermal (power-law) distributions only in parallel shocks, where the magnetic field is nearly parallel to the \identityone{shock normal}. Once the obliquity of magnetic field increases, particle acceleration is inhibited.


Recently, a potential solution to this issue has been proposed. Efficient particle acceleration can be recovered for the case of shocks corrugated (rippled) by upstream inhomogeneities \identityone{\citep[][]{2022Demidem}}. 
Such inhomogeneities may be produced by inhomogeneous magnetic field configurations \citep[e.g.,][]{CeruttiGiacinti2020}, or as a result of operating kinetic turbulence \citep[e.g.,][]{Zhdankin2017}. Interestingly, they were also seen to self-consistently arise in kinetic electron-ion simulations \citep{Ligorini2021}.
These findings can recover internal shocks as an efficient mechanism to \identityone{produce nonthermal particles in} relativistic outflows.

\identityone{The structure behind (downstream of) shocks corrugated by upstream/downstream perturbations is much more complex than for planar shocks in homogeneous media. \cite{Demidem2018} described these perturbations in terms of eigenmodes of linear MHD (i.e., entropy, Alfv\'en, and magnetosonic waves) and studied the response of a fast perpendicular shock to each mode, one at a time. Their MHD simulations illustrate how an incoming upstream plane wave induces shock corrugation and is transmitted into a collection of modes in the downstream (see, e.g., their first figure or, for corrugation induced by the reflection of downstream perturbations, Fig.~6 of \citealt{2012Lyutikov}). \cite{2022Demidem} used a kinetic approach (enabling study of particle energization from first principles) and investigated possible implications for particle acceleration in relativistic magnetized perpendicular shocks. While the process is inhibited for shocks in initially homogeneous media, they find that shock rippling and downstream turbulence induced by (upstream) harmonic density perturbations can lead to generation of energetic particles.} 

\identityone{Inspired by these results, the present work proposes to analyze the properties of a collision of two-dimensional (2D) corrugated shells and how its observational signatures may differ from the idealized (one-dimensional, or, 1D) case of flat shells. }
We perform \identityone{relativistic} magnetohydrodynamic (MHD) simulations of colliding shells, \identityone{corrugated by ambient density or pressure inhomogeneities}, and follow the evolution of density, magnetic field, and energy dissipation (as probed by synchrotron emission).
We find that the quantities are moderately affected by the 2D-structure of the shocks, relative to the 1D case with flat colliding shells.
Hence, we conclude that the previous predictions of analytical models and hydrodynamic simulations hold for the case of acceleration-efficient corrugated shocks.

    \section{Methods}\label{sect:mhd:methods}
    
        The one-dimensional behaviour of colliding shells, including the effects of shell magnetization, initial internal energy, and other parameters, have been investigated in exquisite detail by previous works \citep[e.g.,][]{2001KobayashiSari, 2004FanWeiZhang, Kino2004, Mimica2004, Mimica2005, 2007Mimica, 2010Jamil, 2010MimicaAloy, 2011JoshiBottcher, 2012Mimica, 2017Peer, 2020Rudolph}. Thus, here we treat the one-dimensional behavior of the shells as well established, and instead focus on the difference between a 1D shell collision and the \textit{matching} (see Sect.~\ref{sect:corr_methods}) 2D corrugated case.
        
        In each (1D/2D) case, we simulate a collision of \identityone{two identical dense shells}, which are magnetized uniformly, with magnetic field lines being perpendicular to their motion (corresponding to the so-called perpendicular shock configuration), and moving through a low-density, unmagnetized ambient medium. We simulate the encounter in the zero-momentum frame of the system. For each test case, we run two simulations: a one-dimensional one (henceforth 1D) with no shell structure perpendicular to the shell motion (Sect.~\ref{sect:1Dsetup}), and a two-dimensional simulation (2D) with shell surface corrugation (Sect.~\ref{sect:corr_methods}).
        
        \subsection{Numerical setup}
        
        We use the grid-based astrophysical MHD code \texttt{Athena~4.2}\footnote{\texttt{Athena~4.2} is publicly available at \url{https://github.com/PrincetonUniversity/Athena-Cversion}. A fork of this repository augmented to run and process the simulations presented in this paper is available at \url{https://github.com/ppjanka/Corrugated_internal_shocks}.} \citep{2008Stone}, a Godunov code using constrained transport (CT) method to enforce the zero-divergence condition on magnetic fields. 
        
        Our simulations are performed in special relativistic MHD \citep{2011BeckwithStone_AthenaSR}, in Cartesian coordinates. The boundary conditions in the direction parallel to the shell motion (along the jet) are set as ``free outflow'' (variables at the domain edge are copied to the ghost cells), while those perpendicular to shell motion are periodic. The equation of state is adiabatic with \identityone{the adiabatic index} of $\gamma_{\rm ad}=4/3$. We use 2$^{\rm nd}$-order reconstruction in primitive variables, the van~Leer unsplit time integrator \citep{2009StoneGardiner}, and HLLC \citep{1999ToroHLLC, 2009MignoneHLLC} as our Riemann solver, with $\textrm{CFL}=0.025$.
        
        For our grid resolution, we require that i) for each corrugation method (see Sect.~\ref{sect:corr_methods}) we resolve the corrugation width $w$ of the shock by at least $8$ cells (parallel to the mean shock normal, see Sect.~\ref{sect:diagnostics} and Fig.~\ref{fig:corrugation_aspect_ratios}) and ii) that the resolution is nearly identical in both grid directions (i.e., our cells are nearly square-shaped). This results in the simulation resolution of $8192\times1024$ for a box of \identityone{$40$ light seconds $\times$ $6$ light seconds} (hereafter, lt-sec). For 1D simulations, we use the same resolution along the jet (for fair comparison) and collapse the perpendicular direction to a single cell.

        \subsection{Choice of parameter space}\label{sect:gx_pars}
        
    \begin{table}[]
        \centering
        \begin{tabular}{c|l|l}
            \hline
            \multicolumn{3}{c}{\bf Microphysics} \\
            \hline
            $\gamma_{\rm ad}$ & adiabatic index & $4/3$ \\
            \hline
            \multicolumn{3}{c}{\bf Geometry} \\
            \hline
            \multicolumn{2}{c|}{Box size (lt-sec)} & $40\times 6$ \\
            \multicolumn{2}{c|}{1D box size (cells)} & $8192\times 1$ \\
            \multicolumn{2}{c|}{2D box size (cells)} & $8192\times 1024$ \\
            \multicolumn{2}{c|}{Distance between shells (lt-sec)} & $5$ \\
            \multicolumn{2}{c|}{Shell width (lt-sec)} & $1$ \\
            \hline
            \multicolumn{3}{c}{\bf Shell fluid} \\
            \hline
            $\rho_{\rm sh}$ & Density & $10^{-13}\textrm{ g}/\textrm{cm}^3$\\
            $\sigma$ & Magnetization & $0.1$ \\
            $\beta$ & Plasma $\beta$ & $1$ \\
            $v_{\rm sh}$ & Bulk speed (collision frame) & $0.1c$ \\
            \hline
            \multicolumn{3}{c}{\bf Ambient medium} \\
            \hline
            $\rho_{\rm amb}$ & Density & $10^{-3}\rho_{\rm sh}$ \\
            $P_{\rm amb}$ & Pressure & $10^{-17}(\textrm{g}/\textrm{cm}^3)c^2$ \\
            $B_{\rm amb}$ & Magnetic field strength & 0 \\
            \hline
            \multicolumn{3}{c}{\bf Synchrotron flux} \\
            \hline
            $D$ & Distance & $8$ kpc \\
            $R$ & Size of the emitting region & $6\times10^8$ cm \\
            & (geometrical depth) & \\
            $\Gamma_{\rm j}$ & Jet bulk Lorentz factor & $2$ \\
            $i$ & Jet axis inclination &
            $60\degree$ \\
            & & or $30\degree\sim 1/\Gamma_{\rm j}$\\
            \hline
        \end{tabular}
        \caption{Parameters of our model. Where applicable, initial values are given -- note that these may be altered by the corrugation procedure (both for 1D and 2D runs; as explained in Sect.~\ref{sect:corr_methods}). For synchrotron flux calculations, we use microquasar-relevant values as our fiducial model; we also discuss the application of our results to blazars in text. $c$ denotes the speed of light.}
        \label{tab:parameters}
    \end{table}

To make our calculations more readable, we set our problem parameters and units to match the environment of an existing source (at least to an order of magnitude) and list them in Table~\ref{tab:parameters}. While our calculations are relevant for any relativistic jet where emission is synchrotron-dominated, we provide numerical values for the microquasar \gx. We briefly discuss a comparison with blazars, another prime candidate for application of the internal shock model, at the end of this section.

\gx\ is a well-observed low-mass black hole X-ray binary, with a short duty cycle of X-ray outbursts of $\sim2-3$ years \citep[see, e.g.,][]{2013Corbel, 2019_Connors}. While its orbital period can be measured precisely at $1.76$~days \citep{2003Hynes, 2006Levine, Heida2017}, the distance, compact object mass (and nature), and inclination are difficult to specify from observations. \citet{Heida2017} constrain these to the ranges of $D \gtrsim 5$~kpc, $M\in(2.3,9.5)M_{\odot}$, and $i\in(37\degree,78\degree)$, respectively. For the purpose of this work, we will adopt the values of $D\sim8$~kpc and $M\sim10$~$M_{\odot}$. We also adopt a jet opening angle of $\theta_{\rm j}\sim2.3\degree$ and jet bulk Lorentz factor of $\Gamma_{\rm j}\sim2$ (as assumed by \citealt{2018_Malzac}\identityone{, a value typical of microquasar jets, e.g., \citealt{Tetarenko2019,Zdziarski2022}}). We performed our calculations for two different inclinations: $i=30\degree \sim 1/\Gamma_{\rm j}$ (a head-on jet, corresponding to a blazar-like scenario) and $i=60\degree$ (a jet seen from its side, more typical of X-ray binary jets). However, we find that the relative differences between corrugated and non-corrugated runs (on which we base our analysis, see Sect.~\ref{sect:diagnostics}) are very weakly dependent on inclination. Thus, we only report the results for $i=60\degree$ in this manuscript.

If we scale our simulations to microquasar conditions, our region of interest would be radiating in the IR. Based on \cite{2018_Malzac}, this corresponds to $z\sim10^3-10^6r_{\rm g}$ (see their fig. 4), where $r_{\rm g}=1.5\times10^{6}$~cm for a BH of $10M_{\odot}$. \cite{2011_Shidatsu} estimated the jet IR-emitting region size at $R\sim6\times10^8$~cm. Assuming $i=2.3\degree$, this gives $z\sim 1.5\times10^{10}\textrm{ cm}\sim10^4r_{\rm g}$, which we adopt here.
  
In continuous jet models, applicable up to the radio-emitting regions far in the jet, it is often assumed that the high-energy particle and magnetic field energy densities are in equipartition (e.g., \citealt{2005Hirotani}, \citealt{2012ZdziarskiLubinskiSikora}, \citealt{2015Drappeau}; see, however, \citealt{2015ZdziarskiSikoraPjankaTchekhovskoy}). For simplicity, we also follow this assumption here. 
To obtain the remaining plasma parameters, we use existing theoretical models of the binary from the literature to provide lower and upper constraints. We show detailed derivations and the resulting ranges in Appendix~\ref{sect:gx_pars_derivation}. Informed by these results, we adopt the rest-frame fluid density of $\rho \sim 10^{-13}\textrm{ g}/\textrm{cm}^3$ and magnetization of $\sigma \sim 0.1$.

As noted above, while we report numerical values for a microquasar, our models can also be interpreted in terms of blazar sources. This would require a number of changes in Table~\ref{tab:parameters}, most of which would result in similar enhancements or decreases of synchrotron emission for both the 1D and 2D cases (i.e., change of normalization). We are, however, only interested in the difference between the paired 1D/2D models. We find this difference to be mainly influenced by two parameters. Magnetization $\sigma$ contains information about how dominant magnetic fields are in the dynamics of the problem. For large $\sigma$, magnetic fields are able to prevent corrugation, leading to more similar 1D and 2D results. Fortunately, magnetization of the ballistic parts of a blazar jet can be estimated to $\sigma_{\infty}\sim0.1-0.01$ \citep[see][and references therein]{2009Tchekhovskoy, 2015ZdziarskiSikoraPjankaTchekhovskoy}, so conditions in blazar jets are similar to the ones presented here in that respect. The second parameter of relevance is the bulk Lorentz factor of the jet $\Gamma_{\rm j}$. Faster jets likely contain a broader distribution of shell speeds, with high-speed collisions (large $v_{\rm sh}$) more likely. In the models presented in our work, collisions are driven by adiabatic expansion of the shells into the ambient medium between them. At faster relative speeds, this expansion may become less relevant to the collision outcome. If shell corrugation is caused by pressure imbalance within the shell (see our pressure-corrugated runs, Sect.~\ref{sect:corr_methods}), this will result in flatter shocks, and more similar 1D and 2D runs (i.e., less effect of corrugation in the synchrotron emission).
        
        \subsection{1D initial conditions}\label{sect:1Dsetup}
        
        \begin{figure*}
            \centering
            \includegraphics[scale=0.5]{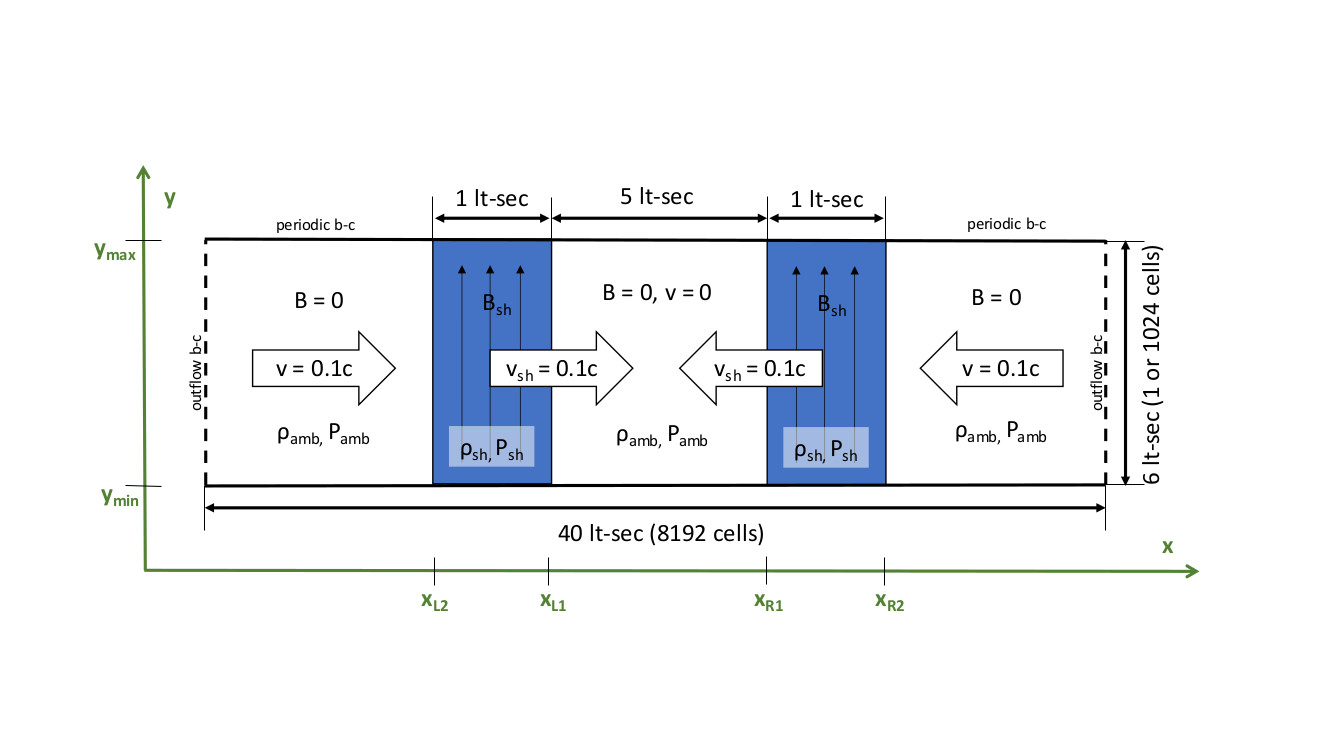}
            \caption{A diagram of our initial conditions and numerical setup (not to scale) before corrugation is applied.}
            \label{fig:initCond_diagram}
        \end{figure*}
        
        Throughout the paper, the fluid quantities \identityone{(density $\rho$ and thermal pressure $P$)} are given in the fluid rest-frame, while the velocity $v$, Lorentz factor $\Gamma$, and magnetic field $B$ are evaluated in the simulation frame (zero-momentum frame of the shell collision). 
        We denote the coordinate along the jet, parallel to the movement of the shells, as $x$ and the perpendicular one as $y$.
        
        The initial conditions for non-corrugated (1D) runs are depicted in Fig.~\ref{fig:initCond_diagram}. The domain is composed of 5 regions delimited by the shell edges: $x_{\rm L1}$, $x_{\rm L2}$ for the left shell and $x_{\rm R1}$, $x_{\rm R2}$ for the right shell, with the subscript ``1'' closer to the center of the domain and ``2'' further away. Note that our setup is symmetric with respect to the plane of the (eventual) shell collision ($x=0$, a vertical line at the center of our domain). Within the shells, density is $\rho_{\rm sh}$, pressure $P_{\rm sh}$, velocity $v_{\rm sh}$ is directed towards the center of the domain, and the magnetic field is set to $B_{\rm sh}\vec{e}_y$. Outside the shells, density equals $\rho_{\rm amb}$, pressure $P_{\rm amb}$ and magnetic field is 0. 
        The velocity field in the ambient medium is initialized according to:
        \beq
        \vec{v} = \left\{\begin{array}{cc}
            v_{\rm sh}\vec{e}_x ,& x \le x_{\rm L1}, \\
            0 ,& x \in (x_{\rm L1}, x_{\rm R1}), \\
            -v_{\rm sh}\vec{e}_x ,& x \ge x_{\rm L1}. \\
        \end{array}\right.
        \eeq
        
        Within the shells, the magnetic field strength $B_{\rm sh}$ and pressure $P_{\rm sh}$ are set through the plasma $\beta$ and $\sigma_B$ parameters:
        \beq B_{\rm sh} = \sqrt{8\pi\rho_{\rm sh}\sigma_{B\rm ,sh}}\Gamma_{\rm sh}c, \eeq
        \beq P_{\rm sh} = \beta\frac{B_{\rm sh}^2}{8\pi\Gamma_{\rm sh}^2}. \eeq
        
        Numerically, we set our initial conditions to $\rho_{\rm sh} = 10^{-13}\textrm{ g}/\textrm{cm}^3$, $\beta = 1$, $\sigma = 0.1$ (cf., Sect.~\ref{sect:gx_pars} and Table~\ref{tab:parameters}), \identityone{$v_{\rm sh} = 0.1c$ (where $c$ denotes the speed of light)}.
        
        \subsection{Shock corrugation methods}\label{sect:corr_methods}
        
        To produce shock corrugation, we modulate values of parameters from the previous paragraph in the direction perpendicular to shell motion. In studies of corrugated shocks, various (typically dynamical) methods are used to induce corrugation, or bending, of the shock front \citep[see, e.g.,][]{Demidem2018}. We investigate two possibilities: density and pressure modulation. \identityone{In real astrophysical sources, such shock-rippling perturbations can follow very complex spatial dependencies. However, they can always be decomposed into harmonic components. Thus, as a first step, we use sinusoidal modulation to represent single harmonics of such shock-rippling density or pressure perturbations in the fluid, in a fashion similar to that of \cite{Demidem2018,2022Demidem}.}
        
        An important aspect of each implementation lies in ensuring that each 1D/2D pair of simulations (with matching global parameters) is only distinguished by the 2D structure of the latter, and not by, e.g., the initial total synchrotron flux. 
        As a result, each modulation in the initial conditions creating a 2D (corrugated) simulation may also include a change in the parameters of the corresponding 1D case (i.e., modify its parameters: $\rho_{\rm amb}$, $\rho_{\rm sh}$, $P_{\rm amb}$, etc.). This is done to ensure that the initial lab-frame \identityone{total energy} (integrated over the simulation domain),
        \beq \epsilon_{\rm tot} = \int \left[ \Gamma^2 \left( \rho c^2 + \frac{\gamma_{\rm ad}}{\gamma_{\rm ad}-1}P \right) - P - \identityone{\Gamma}\rho c^2 \right] dV, \label{eq:conserve_int_energy} \eeq
        shell momentum,
        \beq p_{\rm sh,tot} = \int_{\rm sh} \left[ \Gamma^2 \left(\rho c^2 + \frac{\gamma_{\rm ad}}{\gamma_{\rm ad}-1}P \right) \frac{v}{\identityone{c^2}} \right] dV, \label{eq:conserve_momentum} \eeq
        and synchrotron power (see Sect.~\ref{sect:diagnostics}),
        \beq F_{\rm syn,tot} \appropto B^{\frac{p+5}{2}}, \label{eq:conserve_syn} \eeq
        of paired 1D/2D runs are identical (and thus, the runs are comparable). In the equations above, $\Gamma$ is the local fluid Lorentz factor (in the simulation frame), $\rho$ -- the local fluid-frame mass density, $P$ -- fluid-frame thermal pressure, and $B$ is the simulation-frame magnetic field strength. Conservation of quantities (\ref{eq:conserve_int_energy})-(\ref{eq:conserve_syn}) can be achieved by ensuring that the averages of $\rho$, $P$, and $B$ over the $y$ coordinate match between corresponding 1D/2D cases.
        
        \subsubsection{Density modulation}
        
        In our first method of inducing shell corrugation, that we will refer to as ``density corrugation'', we change the density of ambient medium between shells to exhibit ``ridges'' (sinusoidal density modulation oblique with respect to the shell surface) that bend the surface of each shell as it sweeps their mass. To ensure that the paired 1D run has initially identical global properties (\ref{eq:conserve_int_energy})-(\ref{eq:conserve_syn}), we also increase the ambient medium density $\rho_{\rm amb}$ to match the mean density in the ``ridged'' zone in 2D run. In other words, for the 2D case we set:
        \beq \rho_{\rm 2D} = \left\{\begin{array}{ll}
            \xi\rho_{\rm amb} ,& x < x_{\rm L2}, \\
            \rho_{\rm sh} ,& x \in [x_{\rm L2}, x_{\rm L1}], \\
            \begin{array}{rl}
                \rho_{\rm amb}+\frac{1}{2}&(A_{\rho}\rho_{\rm sh}-\rho_{\rm amb}) \\
                              &\times \mathcal{M}_{\rho}(x,y)
            \end{array} ,& x \in ]x_{\rm L1}, x_{\rm R1}[, \\
            \rho_{\rm sh} ,& x \in [x_{\rm R1}, x_{\rm R2}], \\
            \xi\rho_{\rm amb},& x > x_{\rm R2},
        \end{array}\right.\eeq
        where $A_{\rho}$ is the corrugation amplitude,  $\displaystyle \xi = \frac{1}{2}\left(1+A_{\rho}\frac{\rho_{\rm sh}}{\rho_{\rm amb}}\right)$, and
        \begin{multline}
            \mathcal{M}_{\rho}(x,y) = \cos\left(2\pi(n_x \frac{x-x_{\rm L1}}{x_{\rm R1}-x_{\rm L1}}\right. \\
            \left. + n_y \frac{y-y_{\rm min}}{y_{\rm max}-y_{\rm min}}) \right) +1,
        \label{eq:corrugation_rho_modulation}\end{multline}
        with $(n_x,n_y)=(1,2)$. For the 1D case (to conserve quantities in Eqs. \ref{eq:conserve_int_energy} and \ref{eq:conserve_momentum}), we set
        \beq \rho_{\rm 1D} = \left\{\begin{array}{ll}
            \xi\rho_{\rm amb} ,& x < x_{\rm L2}, \\
            \rho_{\rm sh} ,& x \in [x_{\rm L2}, x_{\rm L1}], \\
            \xi\rho_{\rm amb} ,& x \in ]x_{\rm L1}, x_{\rm R1}[, \\
            \rho_{\rm sh} ,& x \in [x_{\rm R1}, x_{\rm R2}], \\
            \xi\rho_{\rm amb} ,& x > x_{\rm R2}.
        \end{array}\right.\eeq
        
        Note that we only adjust the ambient medium conditions, and leave the fluid within the shells unchanged. Thus, 1D/2D pairs of runs with different $A_{\rho}$ have different density contrast ratios. While the runs with different $A_{\rho}$ are in effect less easily comparable in absolute terms (and so we only use relative values in all our comparisons), it allows us to focus on the effect of corrugation.
        
        Overall, we perform density-corrugation simulations (1D/2D pairs) for $A_{\rho}$ values of $0.01$, $0.02$, $0.05$, $0.10$, $0.20$, $0.50$, $0.75$, and $1.00$.
        
        \subsubsection{Shell pressure modulation}
        
        In the second method of inducing shell corrugation, henceforth referred to as ``pressure corrugation'', we implement a pressure modulation within the shell.
        \begin{multline}
            P_{\rm sh,2D} = P_{\rm sh}+\frac{1}{2}A_{P}(P_{\rm sh}-P_{\rm amb}) \\
                              \times\cos\left[ 2\pi \left(n_x \frac{x-x_{\rm S2}}{x_{\rm S1}-x_{\rm S2}}
                      + n_y \frac{y-y_{\rm min}}{y_{\rm max}-y_{\rm min}}\right)\right],
        \label{eq:corrugation_press}\end{multline}
        where $S\in\{L,R\}$ is the shell index, $(n_x,n_y)=(1,2)$, and $A_{P}$ is the ``corrugation amplitude''. As the conditions in Eq.~(\ref{eq:corrugation_press}) do not change our original mean internal energy, shell momentum, or synchrotron flux, other initial conditions remain unchanged.
        
        We perform pressure-corrugation simulations (1D/2D pairs) for $A_{P}$ values of $0.10$, $0.20$, $0.50$, $0.75$, and $1.00$.
        
        
        
        
        
        
        \subsection{Diagnostics}\label{sect:diagnostics}
        
        \subsubsection{Measuring corrugation}
        
        \begin{figure}
            \centering
            \includegraphics{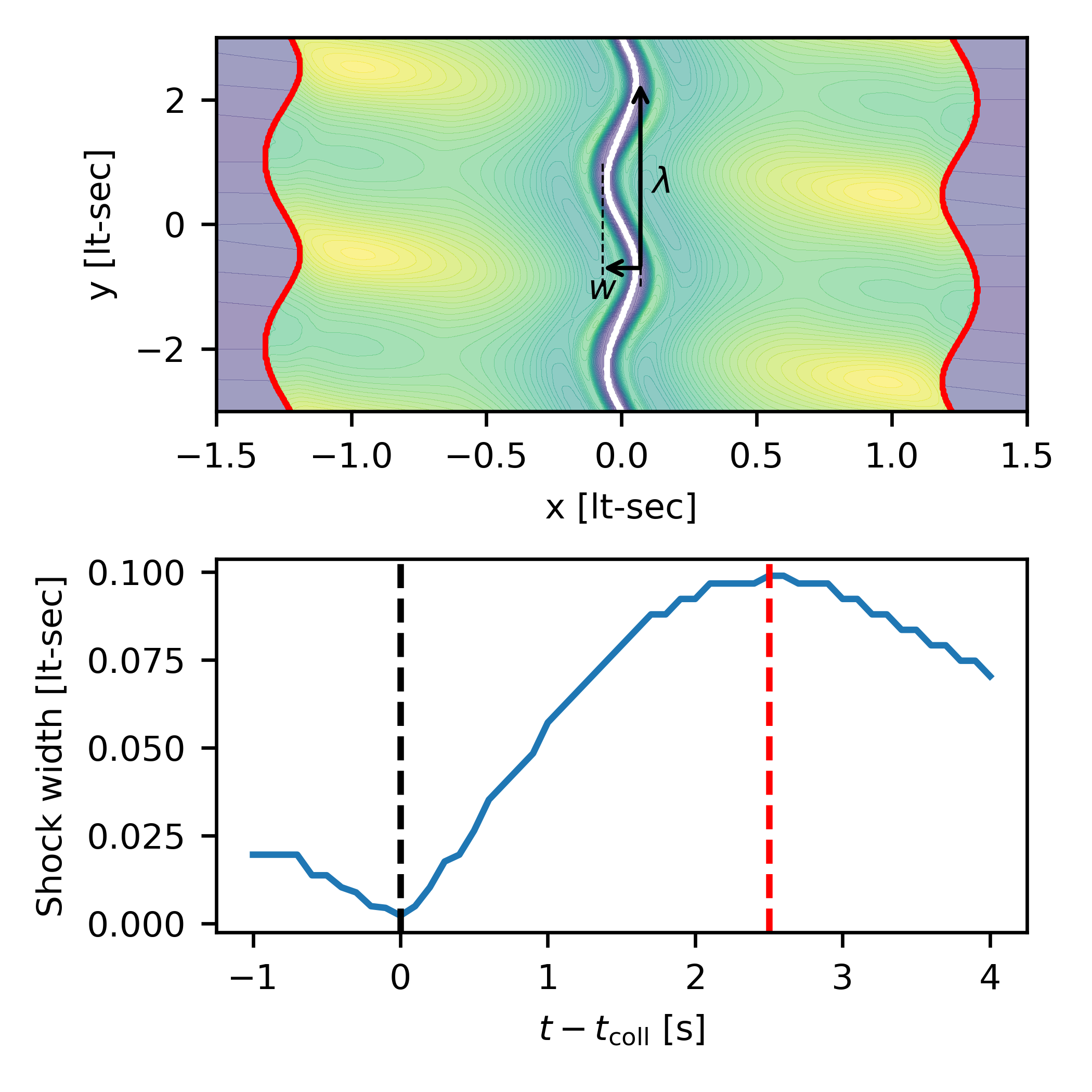}
            \caption{An example measurement of shock corrugation aspect ratio ($a_{\rm sh}=w/\lambda$), for a pressure corrugated case with $A_P=0.2$. Top: a density contour plot with shock detection marked as red lines and aspect ratio measurement depicted with the white curve and annotations (see text). Bottom: shock width $w$ as a function of time (with respect to collision time $t_{\rm coll}$). The dashed black and red lines show the moments of collision and maximal corrugation, respectively. The final shock width $w_{\rm max}$ is measured at the red dashed line.}
            \label{fig:shock_corrugation_measurement}
        \end{figure}
        
        \begin{figure}
            \centering
            \includegraphics{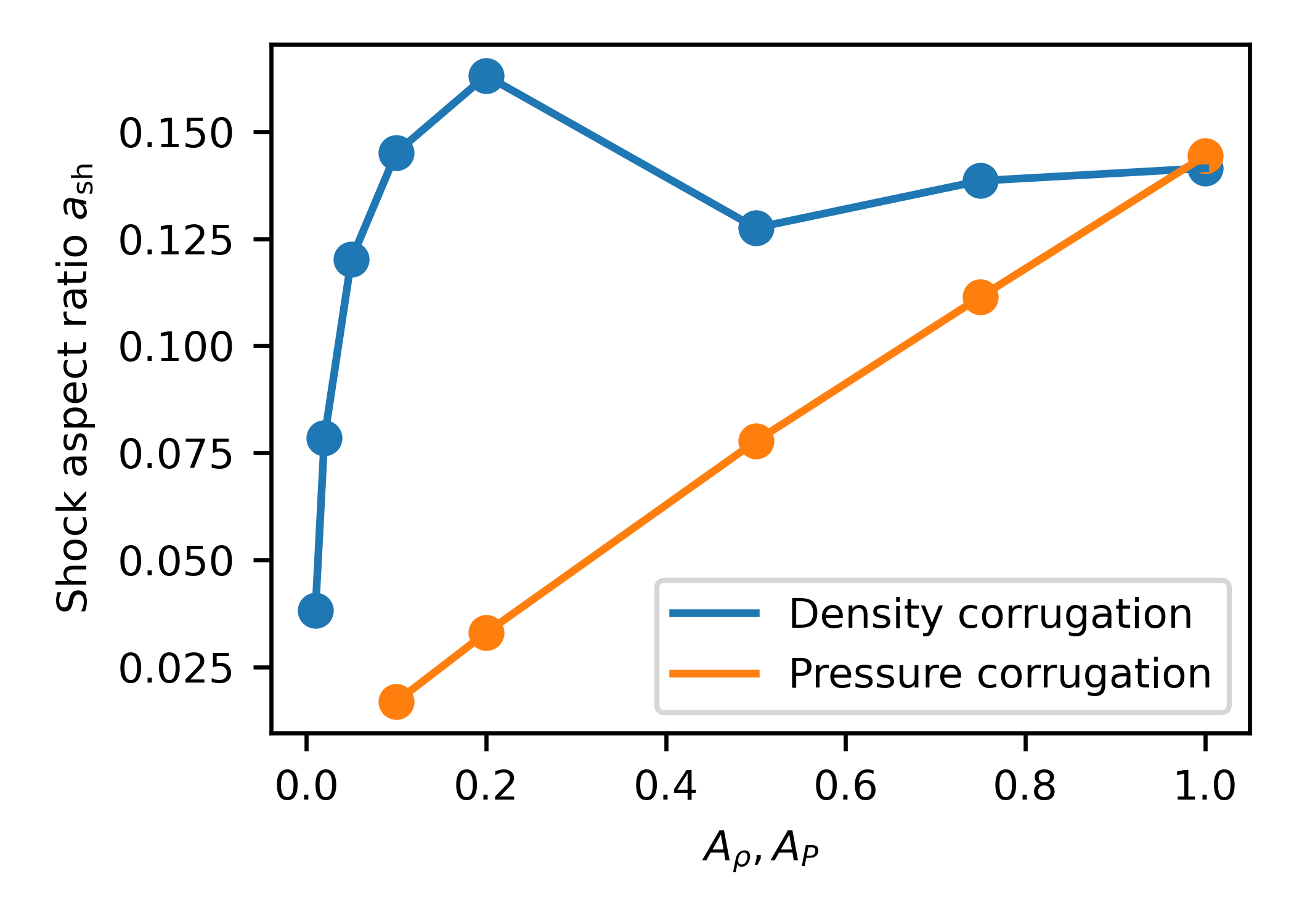}
            \caption{Scaling between the shock aspect ratio and the initial conditions' corrugation parameter for density corrugation (blue) and pressure corrugation (orange).}
            \label{fig:corrugation_aspect_ratios}
        \end{figure}
        
        In our models, shell corrugation is induced in the initial conditions, via the modulation fractions $A_{\rho}$ and $A_{P}$. These parameters tell us little about the eventual shock geometry. Therefore, we explicitly measure the shape of the corrugated shock we obtain in each simulation. As our diagnostic, we use the shock's ``aspect ratio'' (see Fig.~\ref{fig:shock_corrugation_measurement}), the ratio of its width $w$, the distance between maximum corrugation down- and upstream, to its wavelength $\lambda$, set by the parameters $(n_x,n_y)=(1,2)$ in Eqs. (\ref{eq:corrugation_rho_modulation}--\ref{eq:corrugation_press}).
        
        From the 1D version of a given run, we extract the maximal density for each frame, and note the time of maximal compression -- which we consider the ``collision time'' $t_{\rm coll}$. For the 2D version, we inspect the frames between $t_{\rm coll}-t_{\rm cross}$ and $t_{\rm coll}+nt_{\rm cross}$ (with $n=2,4$ for density and pressure corrugation, respectively), where $t_{\rm cross}$ is the light-crossing time of the initial shell width. For each or those frames, we measure the instantaneous shock width:
            \begin{enumerate}
                \item In each frame, the center of the frame is surrounded from both sides by two shocks, either driven by shells propagating through the ambient medium ($t<t_{\rm coll}$), or driven into each shell by the collision ($t>t_{\rm coll}$). We localize these shocks using the velocity gradient. $x$-coordinates of points with $|\partial v_x / \partial x|$ larger than a threshold value are averaged along the $x$ direction separately for the left and right halves of the simulated domain, giving us the precise locations of both shocks at each $y$ (shown by red curves in Fig.~\ref{fig:shock_corrugation_measurement}).
                \item At each $y$, we average the $x$-position of these two shocks horizontally, to obtain the white stripe (along the center) in Fig.~\ref{fig:shock_corrugation_measurement}.
                \item We then collect the $x$ coordinates of all points within the white stripe (one per each $y$ value) and produce their histogram. To discard the outliers, we take the distance between the maxima of this histogram to be the width of the shock in a given frame.
            \end{enumerate}
        
        In all our simulations, we find that the shock width measured this way decreases before the collision, reaches a minimum at $t_{\rm coll}$, then increases and reaches a maximum, after which the \identityone{width continuously decreases as the post-shock regions evolve}. We take the post-collision maximal shock width $w_{\rm max}$ as the shock width for the given 1D/2D configuration (i.e., for the given $A_{\rho}$ or $A_{P}$). The corrugation wavelength is always $\Delta y / 2$ (where $\Delta y$ is the simulated box's height; cf. Eqs. (\ref{eq:corrugation_rho_modulation}) -- (\ref{eq:corrugation_press})), giving us the aspect ratio of:
            \beq a_{\rm sh} = 2w_{\rm max} / \Delta y. \eeq
        We plot the relations between $a_{\rm sh}$ and $A_{\rho}$, $A_{P}$ in Fig.~\ref{fig:corrugation_aspect_ratios}.
        
        \subsubsection{Synchrotron emission}
        
        The main deliverable of internal shock models lies in their accurate prediction of jet emission which, in the leptonic model adopted here, is powered by the non-thermal electron population. Of course, an MHD-only picture does not include non-thermal particles, and thus a number of assumptions must be adopted to extract the desired observables. We select our assumptions in line with typical prescriptions from the literature \citep[cf., e.g.,][]{Spada2001, 2008Graff, BoettcherDermer2010, 2010Jamil, 2014Malzac, 2014RuedaBecerril, 2018_Malzac}. At each location within the jet, the local non-thermal electron energy distribution is assumed to be a power law $dN/d\gamma \propto \gamma^{-p}$ with a slope of $p=2.5$, extending between the Lorentz factors $\gamma_{\rm min}=10$ and $\gamma_{\rm max}=10^6$. Non-thermal electrons are taken to be in energy equipartition with the magnetic field (i.e., with equipartition parameter $\zeta_e=U_e/U_B=1$, where $U_e=\int (\gamma m_ec^2 \frac{dN}{d\gamma})d\gamma$ and $U_B = B^2/(8\pi)$).
        
        We stress that, as a consequence of the assumptions above, the impact of thermal particles on the acceleration and radiative processes is assumed to be negligible -- thermal particles (represented by the mass density in MHD simulations) only affect the fluid dynamics of the model. Particle acceleration is assumed to proceed until there is equipartition between the \textit{non-thermal} particles and the magnetic field. Thus, the number density of non-thermal particles is a direct function of the magnetic field energy density. This is regardless of the MHD mass density, which is assumed to be sufficiently large to provide seed particles for the non-thermal distribution without itself being significantly depleted. In turn, we also assume that synchrotron emission and self-absorbtion are both completely dominated by non-thermal particles -- we caution that this last assumption may not be applicable in some blazars, where self-absorbtion by thermal electrons may be important.
        
        For our results, we report the synchrotron flux per unit jet surface area in the observer's frame \citep[e.g.,][]{RybickiLightman}:
        \beq \frac{dF_{\nu}}{dS} = \frac{\delta^2\Gamma_{\textrm{fl}\rightarrow\textrm{obs}}}{2D^2}f_v\bar I_{\bar\nu}, \label{eq:syn_flux}\eeq
        where $\delta = [\Gamma_{\textrm{fl}\rightarrow\textrm{obs}}(1-\vec{v}_{\textrm{fl}\rightarrow\textrm{obs}}\cdot\vec{n}\identityone{/c})]^{-1}$ is the Doppler factor, $\vec{n}$ is the direction towards the observer, $\Gamma_{\textrm{fl}\rightarrow\textrm{obs}}=1/\sqrt{1-v^2_{\textrm{fl}\rightarrow\textrm{obs}}\identityone{/c^2}}$, $f_{\nu} = 1$ is the volume filling factor of non-thermal electrons, $\nu = \bar\nu\delta$ is the observer frame photon frequency linked to the jet-frame frequency $\bar\nu$, and $\bar I_{\bar\nu}$ is the jet-frame synchrotron intensity. In order to determine the values of the fluid quantities in the observer's frame from their rest-frame values, one needs to account for both the velocity $\vec{v}=(v_x,v_y)$ of the fluid in the simulation frame (zero-momentum frame of the shell collision) and the speed of the simulation frame in the observer frame $\vec{v_{\rm j}} = v_{\rm j}\vec{e}_x$ (bulk velocity of the jet). We do so by performing Lorentz transformations of the fluid velocity:
        \beq \vec{v}_{\textrm{fl}\rightarrow\textrm{obs}} = \frac{v_x + v_{\rm j}}{1+v_xv_{\rm j}\identityone{/c^2}}\vec{e}_x +  \frac{\sqrt{\identityone{c^2}-v_{\rm j}^2}v_y}{\identityone{c^2}+v_xv_{\rm j}}\vec{e}_y. \eeq
        We consider the total IR-opt flux $dF_{\rm syn}/dS$, which we take to correspond to the flux integrated over the frequency range $300$~GHz -- $3$~PHz range. For our assumed electron distribution, the local intensity can be written as \citep[e.g.,][]{RybickiLightman, Beloborodov2000, 2014Malzac}:
        \beq \bar I_{\bar\nu} = \frac{j_{\bar\nu}}{\alpha_{\bar\nu}}\left( 1-e^{-\alpha_{\bar\nu}R} \right), \eeq
        where the synchrotron emission and extinction coefficients are:
        \beq j_{\bar\nu} = K_j\zeta_eB^{\frac{p+5}{2}}\bar\nu^{-\frac{p-1}{2}}, \eeq
        \begin{multline}
            K_j = \frac{\sqrt{3}e^3i_{\gamma}}{16\pi^2m_e^2c^4(p+1)} \\
            \times \Gamma\left(\frac{3p+19}{12}\right)\Gamma\left(\frac{3p-1}{12}\right)\left(\frac{m_ec}{3e}\right)^{-\frac{p-1}{2}},
        \end{multline}
        \beq \alpha_{\bar\nu} = K_{\alpha}\zeta_eB^{\frac{p}{2}+3}\bar\nu^{-\frac{p+4}{2}} \eeq
        \begin{multline}
            K_{\alpha} = \frac{\sqrt{3}e^3i_{\gamma}}{64\pi^2m_e^3c^4}\left(\frac{3e}{2\pi m_ec}\right)^{\frac{p}{2}} \\
            \times\Gamma\left(\frac{3p+2}{12}\right)\Gamma\left(\frac{3p+22}{12}\right),
        \end{multline}
        where $m_e$ and $e$ are the electron mass and unit charge, respectively, and, for convenience, we define \citep[following][]{2014Malzac}:
        \beq i_{\gamma}^{-1} = \bar\gamma_e\int_{\gamma_{\rm min}}^{\gamma_{\rm max}}\gamma^{-p}d\gamma \eeq
        with $\bar\gamma_e$ being the mean electron Lorentz factor. For the optical depth $\alpha_{\bar\nu}R$, we select $R=6\times10^8$~cm to match the size of the IR-emitting core of the jet \citep{2011_Shidatsu}. In other words, we report our simulated box's flux as if it were located along the jet axis and observed with a line of sight at an angle of $1/\Gamma_{\rm j}$ to the jet axis in the observer frame (i.e., perpendicular to the jet axis in the comoving frame).

    \section{Results}\label{sect:mhd:results}
    
\begin{figure*}
        \centering
        \includegraphics{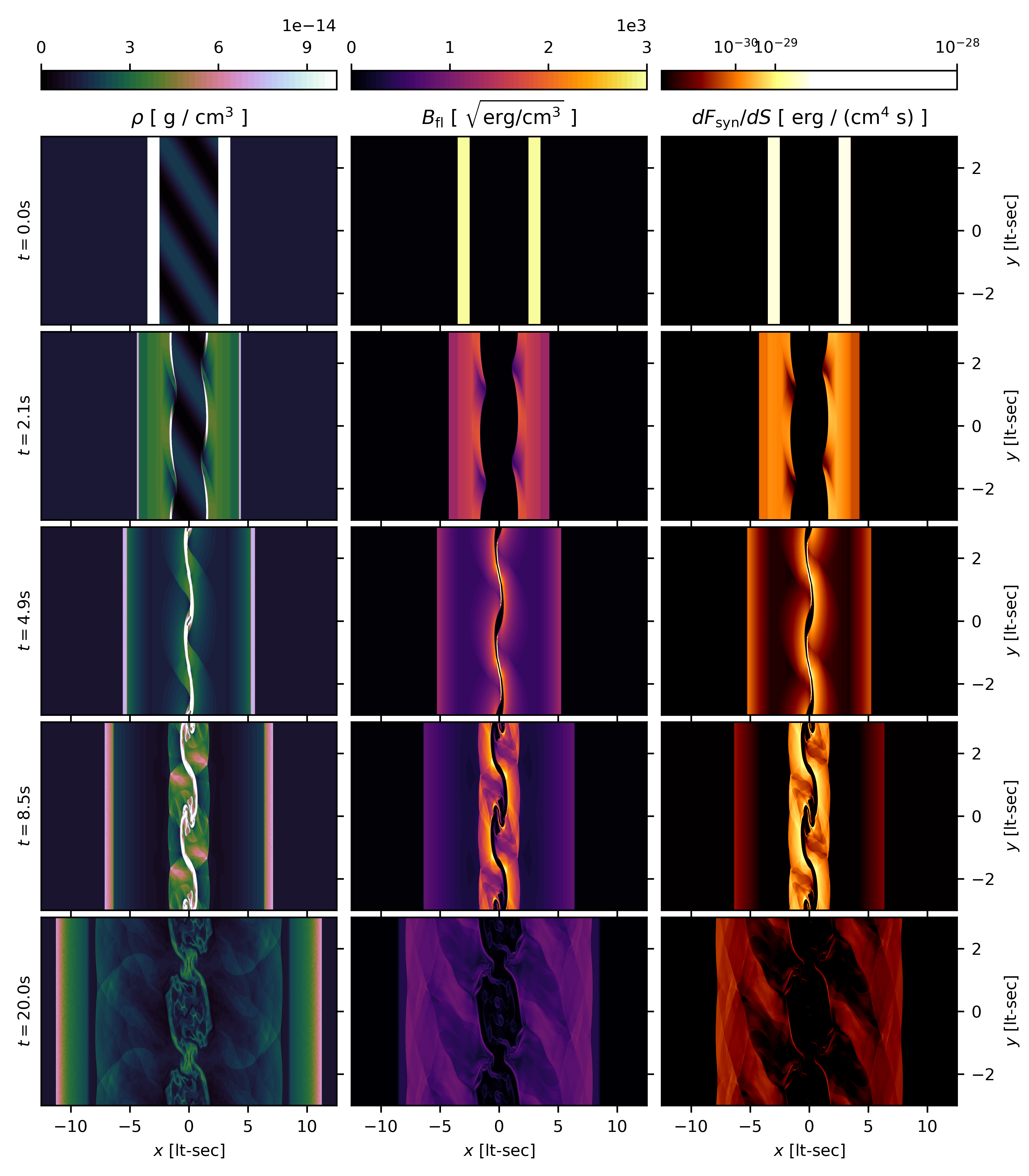}
        \caption{A representative selection of snapshots of shell collision evolution, taken from the density-corrugated (2D) simulation with $A_{\rho}=0.2$ observed at the inclination of $i=60\degree$. Left column: fluid-frame density, middle column: fluid-frame magnetic field strength, right column: observer-frame synchrotron flux per unit emitting surface (log-scale). Each row corresponds to a different time as annotated on the left (cf. Fig.~\ref{fig:dFsyn_vs_time}). The linear and non-linear evolution of shell corrugation can be seen with the progression of time. \identitytwo{Note that, for the density corrugation case shown here, only the inner shell edges (ones facing the other shell) are corrugated, while the outer shell edges remain flat as they propagate outwards into the ambient medium.}}
        \label{fig:snapshots}
    \end{figure*}

    A typical sequence of events during one of the 2D simulations is shown in Fig.~\ref{fig:snapshots}. As the shells are over-pressurized with respect to their surrounding ambient medium, the dynamics of the system is driven by adiabatic expansion of the shells (for a thorough study of the dependence of shell collisions on the initial internal energy see \citealt{2017Peer}). The shells start expanding into the ambient medium with speeds in excess of their bulk velocities and evolve under influence of the corrugation mechanism (see Sect.~\ref{sect:corr_methods}). Their surface ripples in a linear manner, resulting in a near-sinusoidal shape. As the shells collide, two shocks are driven into the shell material (one moving left into the left shell and one moving right into the right one) and a thin line of (unmagnetized) compressed ambient medium is trapped at a contact discontinuity between the two post-shock regions. The post-shock material's density and magnetic field strength are both enhanced, with the latter causing strong synchrotron emission (see Sect.~\ref{sect:resuls:synEnhancement} for further details). Note that, due to our assumption of energy equipartition between the non-thermal particles and the magnetic field, synchrotron emission is completely determined by the latter. As the collision material continues to evolve, it enters a non-linear phase, during which \identitytwo{the central part of the domain becomes disrupted}. In the bottom two rows of Fig.~\ref{fig:snapshots}, development of the Kelvin-Helmholtz instability can be seen at $x\sim0$ to mediate this process. While the box-integrated total synchrotron flux remains high (albeit oscillating with time), the emission becomes spread over the growing post-shock region, and the local flux density dims considerably (as seen in the lower right corner of Fig.~\ref{fig:snapshots}).
    
    
    \subsection{Corrugation-driven synchrotron flux enhancement}\label{sect:resuls:synEnhancement}
        
\begin{figure*}
    \centering
    \includegraphics{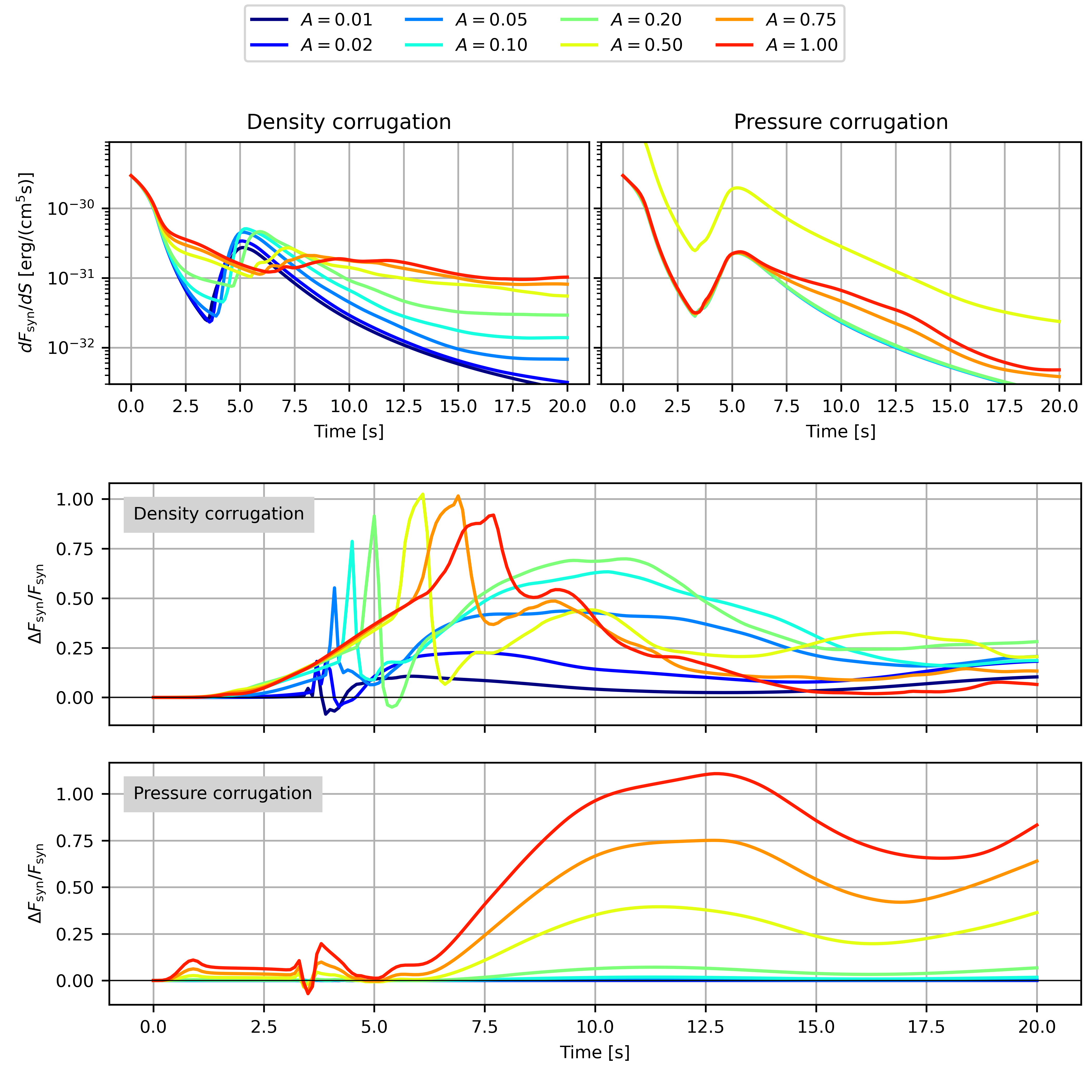}
    \caption{Top row: Light curves (mean synchrotron flux per surface emitting area) for all of our simulations viewed at inclination of $i=60\degree$. Middle and bottom rows: the difference between corrugated (2D) and non-corrugated (1D) matched simulations shown as relative difference in synchrotron flux integrated over the whole simulation domain (i.e., $2(F_{\rm syn-tot, 2D}-F_{\rm syn-tot, 1D})/(F_{\rm syn-tot, 2D}+F_{\rm syn-tot, 1D})$). The curves are color-coded by corrugation amplitudes $A_{\rho}, A_P$ (see Fig.~\ref{fig:corrugation_aspect_ratios} for translation into shock aspect ratios).}
    \label{fig:dFsyn_vs_time}
\end{figure*}
    
    We investigate how corrugation affects synchrotron flux from the shell collision site. Fig.~\ref{fig:dFsyn_vs_time} shows the light curves integrated over our simulation boxes (top row) and the relative difference between synchrotron fluxes of the matched corrugated (2D) and non-corrugated (1D) cases. We notice two effects introduced by shell corrugation.
    
    There is a peak-then-dip pattern in the relative difference curves around collision time ($\sim 3-8$~s). This is caused by the difference in collision time. Due to corrugation, some regions of the 2D shell are travelling ahead of the 1D shell position, and so the shells start colliding earlier in the corrugated (2D) case -- thus giving more emission than the 1D case just before the 1D collision time (when the 2D collision happens), and less emission just afterwards (which is at the 1D collision time, but already after the 2D collision peak).
        
\begin{figure}
    \centering
    \includegraphics{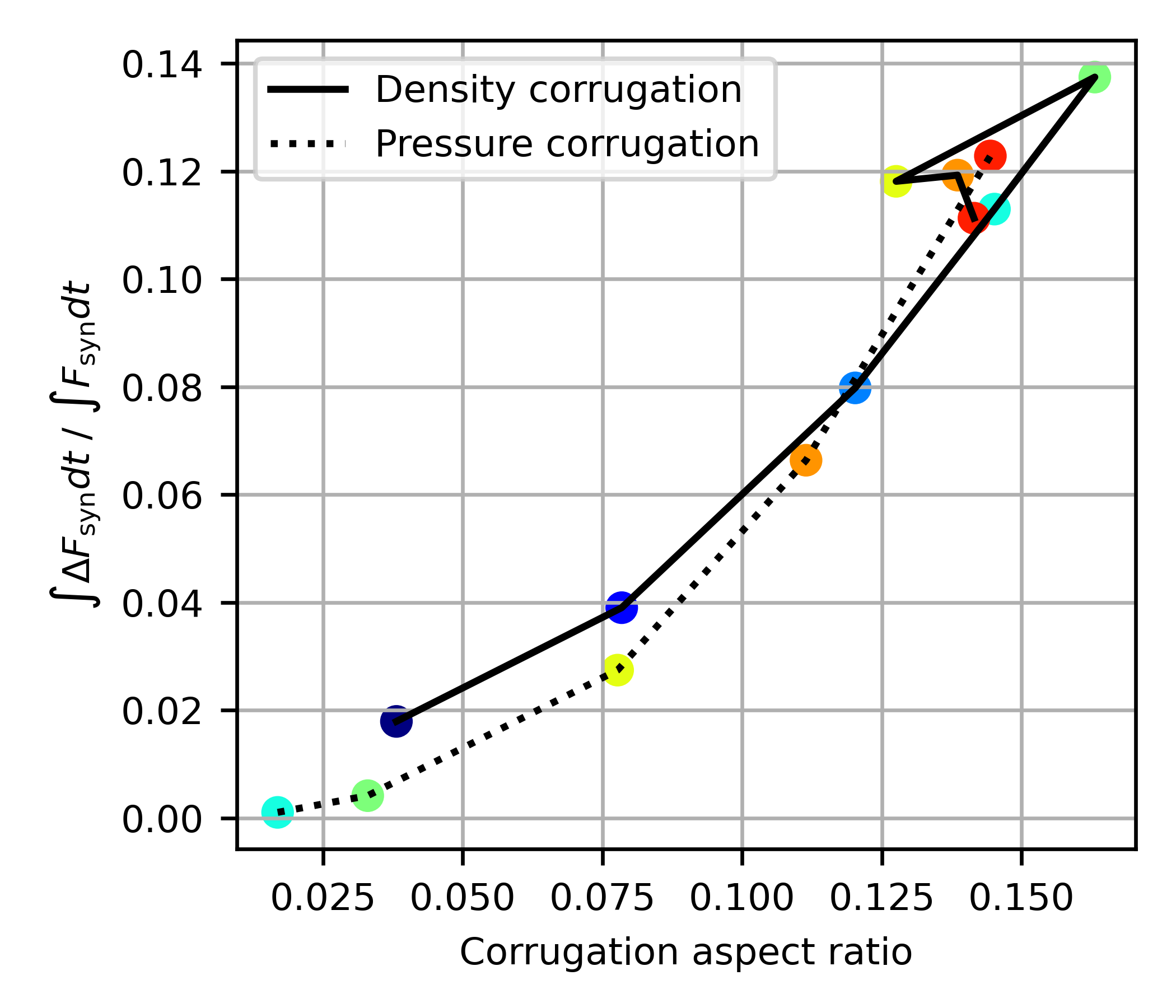}
    \caption{Effect of corrugation on the total synchrotron emission. Dependence of the relative difference in the total synchrotron emission (flux integrated over the simulated domain and simulation time) between the corrugated (2D) and non-corrugated (1D) case, as a function of the shock aspect ratio (see Sect.~\ref{sect:diagnostics} and Fig.~\ref{fig:corrugation_aspect_ratios}). The point colors match the corresponding curves for each run and corrugation method in Fig.~\ref{fig:dFsyn_vs_time}. Note that, as a ratio of synchrotron fluxes, this relation is independent of viewing inclination.}
    \label{fig:dFsyn_total}
\end{figure}
    
    A more interesting result is visible in the long-term behavior of the relative synchrotron flux differences. While the curves oscillate after collision time, there is a clear overall enhancement of the synchrotron flux after shell collision in the 2D case\footnote{Note that the outflow boundary conditions in the direction along the flow remain causally disconnected from the center throughout the simulation (and are therefore insensitive to the collision geometry). Thus, this is a genuine effect of corrugation and not an artifact caused by boundary conditions.} -- especially visible for the pressure-corrugated case (bottom row of Fig.~\ref{fig:dFsyn_vs_time}). To quantify this effect, we integrate the light curves over simulation time, to obtain the total synchrotron emission (per unit surface) for each simulation. We then subtract the 1D from the 2D result and divide by their mean to obtain a total relative enhancement of emission with respect to the 1D case. We plot these differences in Fig.~\ref{fig:dFsyn_total}. We find that the differences can reach $14\%$ in our parameter range. They also scale as a power law with the shock aspect ratio, with a slope of $1.5\pm0.3$ and $2.38\pm0.02$ for density- and pressure-corrugated runs, respectively. Notably, the two curves for density- and pressure-corrugated simulations match almost exactly. Therefore, the observed synchrotron enhancement is independent of the corrugation mechanism and only sensitive to the actual shock geometry.

\begin{figure}
    \centering
    \includegraphics{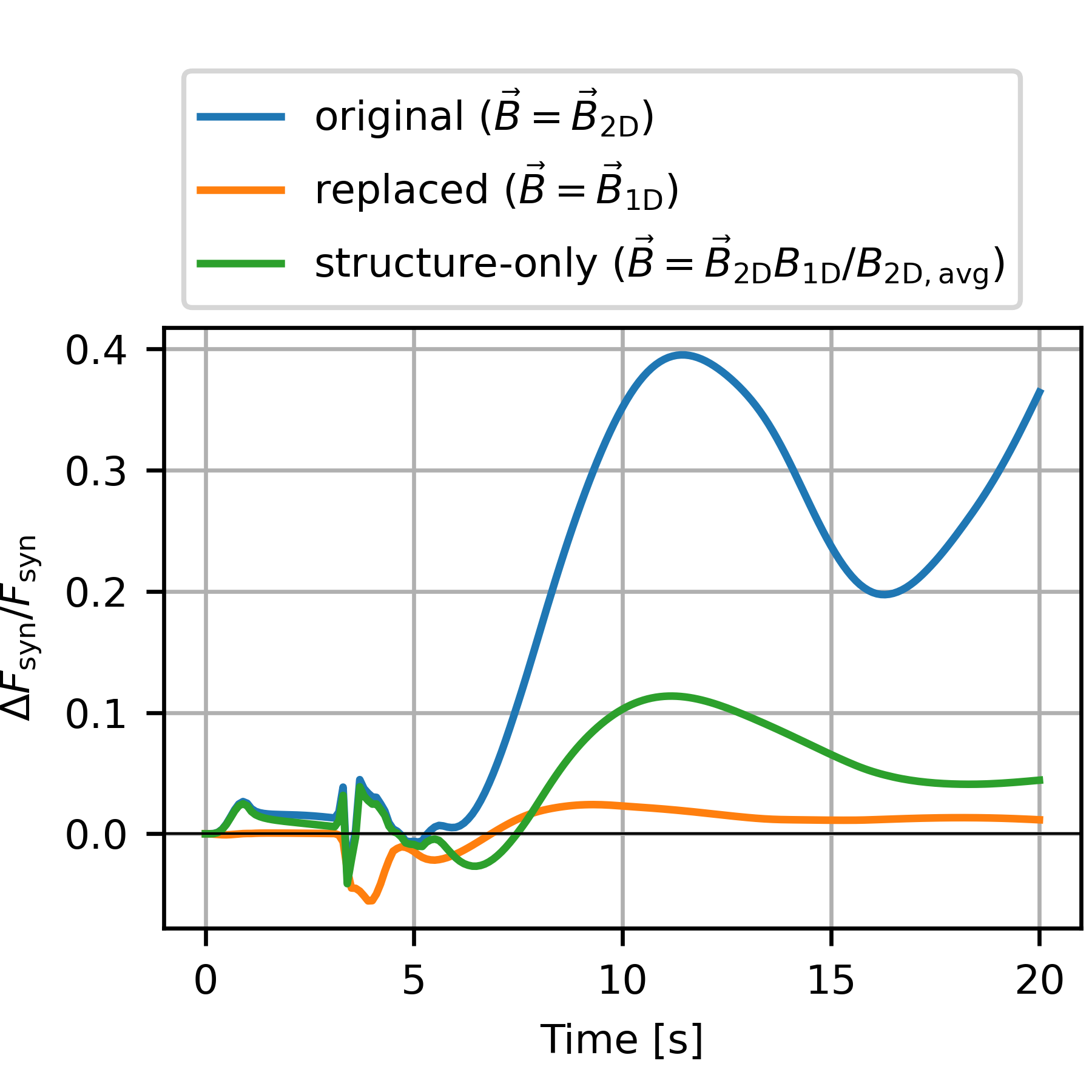}
    \caption{Investigation of the cause of long-term synchrotron emission enhancement. The curves show the relative synchrotron enhancement for the corrugated (2D) simulation with respect to the matched uncorrugated one (1D; cf. the bottom panels of Fig.~\ref{fig:dFsyn_vs_time}) for the pressure corrugated case with $A_P=0.5$. To calculate the emission in the corrugated case, the magnetic field was either kept unchanged (blue curve), replaced with the magnetic field from the uncorrugated (1D) case (orange), or the value was scaled down so that the mean magnetic field strength matches that of the 1D simulation, with the 2D structure left unchanged (green). It can be seen that both the 2D structure and an overall magnetic field enhancement contribute to the enhanced synchrotron emission for the corrugated shell collisions.}
    \label{fig:FsynExp}
\end{figure}

    Under the approximations adopted to calculate the synchrotron flux from the MHD simulation output (namely, equipartition between the nonthermal particle and magnetic energy densities, see Sect.~\ref{sect:mhd:methods}), the synchrotron flux depends only on magnetic field. However, there are two ways in which the magnetic field structure can affect emission. On one hand, an overall increase in the mean magnetic field strength will straightforwardly make the collision brighter. On the other hand, synchrotron emission depends non-linearly on the magnetic field strength (see Eq.~\ref{eq:conserve_syn}). Thus, even if the mean magnetic field strength is identical between 1D and 2D simulations, the presence of 2D structures in the latter can still alter the resulting synchrotron flux. To investigate which of these effects dominates in our case, we perform additional experiments. In a selected 2D run with well-visible synchrotron emission enhancement (pressure-corrugated case with $A_P=0.5$), we replace the magnetic field data with various modifications that deactivate some of the mechanisms above. The results are shown in Fig.~\ref{fig:FsynExp}. The blue curve therein shows the original result, i.e., the 2D simulation with the original 2D magnetic field structure (c.f., Fig.~\ref{fig:dFsyn_vs_time}). The orange curve corresponds to the same simulation where we replaced its original 2D magnetic field with the 1D result (i.e., the 1D magnetic field profile replicated along the $y$ direction to fill the 2D simulation box). As this nullifies both of the mechanisms discussed -- there is no change in mean magnetic field and no 2D structure -- this results in near-zero relative difference\footnote{The remaining difference in flux is due to the 2D velocity structure still present after replacing magnetic field data. The fluid velocity affects transformation of magnetic fields from the observer frame (where the magnetic fields are defined) to the fluid frame (where we calculate synchrotron emission).}. To obtain the green curve, we use the 2D magnetic field result, but scale it so that the mean value matches that of the 1D result. Thus, any enhancement in the mean value is removed, leaving only the influence of dimensionality (2D structure). We see that the resulting curve lies between the orange (no $F_{\rm syn}$ enhancement) and blue (the actual 2D result). Thus, the 2D nature of the magnetic field spatial distribution significantly contributes to the long-term synchrotron emission enhancement in corrugated simulations, but magnetic field amplification (stronger in corrugated than in 1D shocks), has comparable, if not slightly dominant, influence.

   \subsection{Downstream magnetic field structure and polarization properties}\label{sect:resuls:polarization}
   
   \begin{figure}
       \centering
       \includegraphics{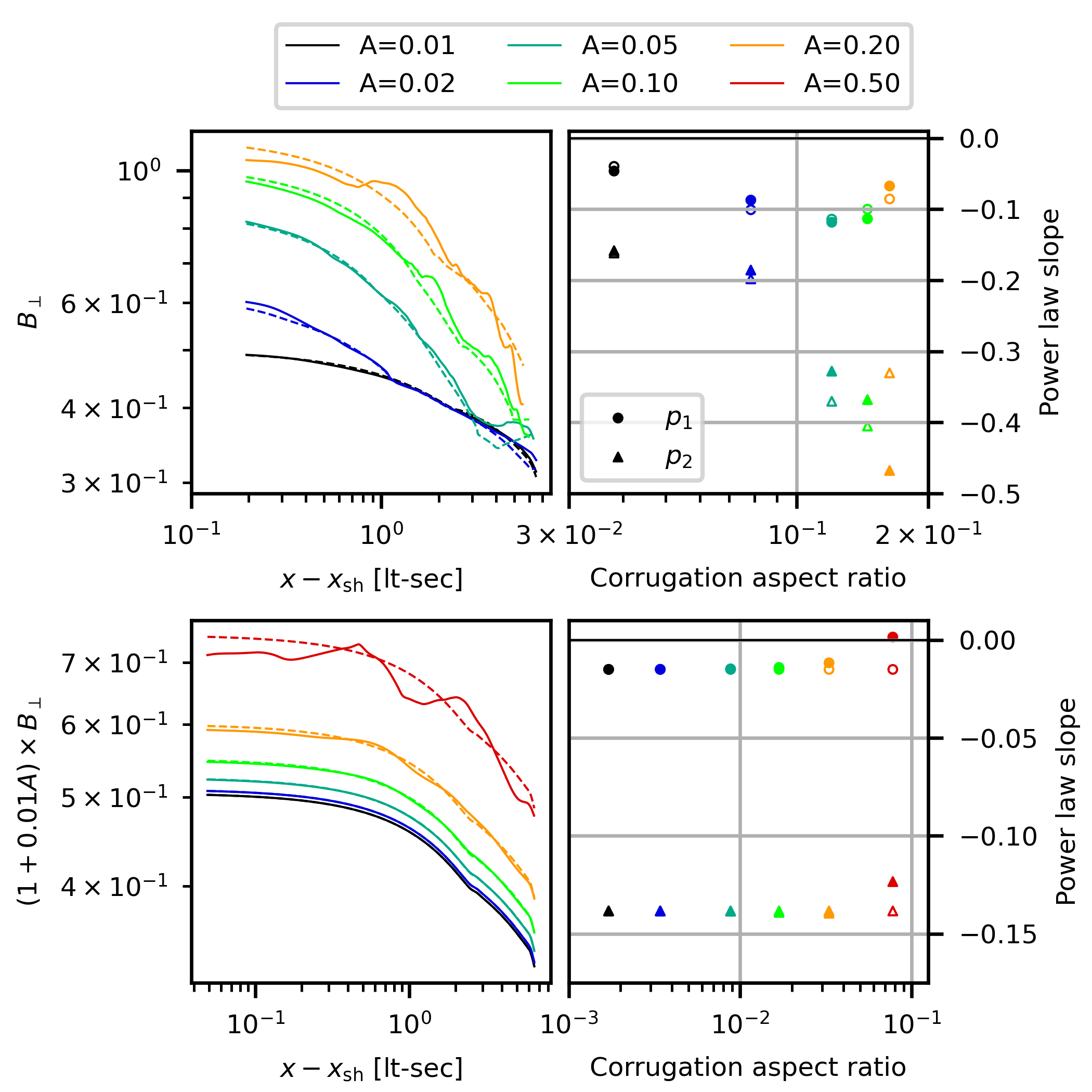}
       \caption{Decay of the magnetic field component perpendicular to the shock normal, $B_{\perp}$. Left: $B_{\perp}$ as a function of the distance from the shock front, $x-x_{\rm sh}$. Solid lines show the vertically-averaged values from the 2D simulations, dashed -- from 1D runs. Right: slopes of the two power laws fit to the measurements ($p_1$ -- close to the shock front, $p_2$ -- further away), as a function of the corrugation aspect ratio. Open and full symbols refer to the fits to 1D and 2D results, respectively. The top row corresponds to density-corrugation simulations, while bottom row plots show pressure-corrugation results. All curves and markers are color-coded by the simulation amplitude, as indicated in the legend.}
       \label{fig:polarization}
   \end{figure}
   
   Corrugation of the shock front affects the structure of magnetic field in the downstream medium (see Fig.~\ref{fig:snapshots} and \citealt{Demidem2018}). Information about the geometry of magnetic field lines has been successfully extracted from observations using polarimetry. In this section, we investigate how the corrugated shell collisions affect the magnetic structure of the downstream and the multiwavelength polarimetric properties. Synchrotron polarization is orthogonal to the direction of magnetic field and is expected to be either parallel or orthogonal to the jet direction for the cases of purely toroidal and poloidal configurations, respectively. In this respect, the ratio of the magnetic field components behind the shock front is of high importance. Previously, the dominance of the parallel field component was assumed at the shock front \citep{2018Tavecchio}, motivated by earlier studies of particle energization possibilities at the shock fronts \citep[][]{SironiSpitkovsky2009, 2013Sironi}. \citet{2018Tavecchio} find that the observed polarization fractions of optical and X-ray emission in BL~Lac jets can be deduced from the slope of the power law fitting the decay of the shock-perpendicular magnetic field $B_{\perp}$ (the component parallel to the shock front, i.e., perpendicular to the shock normal) behind the shock.
   
   In their analysis, \cite{2018Tavecchio} assume the pre-shock magnetic field to be composed of a shock-parallel component (along the flow direction / the shock normal) following a well-defined profile, and a randomly oriented shock-perpendicular component. In this work, we consider a homogeneous shock-perpendicular configuration. However, the post-shock magnetic field strength is always locally dominated by the compressed shock-perpendicular component (randomly oriented in the case of \citealt{2018Tavecchio}, see their fig.~3). Thus, the post-shock region of our model can be viewed as a small region of the post-shock domain of \cite{2018Tavecchio}, where the magnetic field can be treated as homogeneous.
   
   Given these considerations, we can measure the decay slope of $B_{\perp}$ in our simulations and discuss what it implies for the measured polarization in light of the study by \cite{2018Tavecchio}. After finding shock positions using the local maxima of $|\partial B_{\perp} / \partial x|$, we plot the $y$-averaged $B_{\perp}$ as a function of distance from the shock in Fig.~\ref{fig:polarization}. Since each simulation contains two shocks (in the left and the right shell) the average between the two is used.
   
   We find that $B_{\perp}$ remains constant close to the shock and then drops, approximately as a power law. This structure indicates that there is a rarefaction trailing each shock, with a near-homogeneous fluid between them. Due to the fact that the ambient medium between the shells (through which the rarefaction travels) is adjusted between runs for density corrugation (see Sect.~\ref{sect:corr_methods}), we observe much greater diversity of curve shapes in that case (top row of Fig.~\ref{fig:polarization}), while runs with pressure corrugation (bottom row of Fig.~\ref{fig:polarization}) all behave similarly. In each case, the behavior of the matching 1D and 2D runs is very similar, showing that the quantities linear in $B$, such as the mean $B_{\perp}$ considered here, are well approximated by the solution to the 1D Riemann problem.
   
   A (hard-)broken power law fit to $B_{\perp}(x-x_{\rm sh})$ yields the results presented in the right column of Fig.~\ref{fig:polarization}. We see that the post-shock power law slope remains within the range $[-0.50,-0.15]$ for density-corrugated simulations and is well constrained at $\sim -0.14$ for the pressure-corrugated ones. \cite{2018Tavecchio} find that the slope (in their notation, $-m$, see their fig.~4) of $\sim -0.4$ results in high polarization fractions of both the optical and X-ray band, with the latter somewhat higher than the former, $\gtrsim 45\%$ vs $\sim 33\%$. We caution that these specific values would only apply to the case of a single shell collision. Once a lot of events overlap (with some collisions likely off-center, at different inclinations to the jet axis), the expected total polarization fraction of jet emission should be lower, and could be comparable to X-ray polarization recently detected by \texttt{IXPE} \citep{Liodakis2022}. Nonetheless, unless another source of polarized radiation contributes to the observed bands, the X-ray emission should remain more strongly polarized than the optical band.

\section{Discussion and conclusions}\label{sect:conclusions}

In this paper, we revise the viability of internal shock collisions to power the emission observed from relativistic jets. Given sufficient presence of high-energy particles, synchrotron emission in these environments is largely dictated by the strength and orientation of magnetic fields (see discussion in Sect.~\ref{sect:mhd:methods}). At the base of the jet, these may be randomly, or even poloidally, oriented \citep{2003Vlahakis, 2009Komissarov, 2022Zdziarski}. However, the magnetic fields parallel ($B_{||}$) and perpendicular ($B_{\perp}$) to the jet axis evolve differently with the distance from the central engine \citep{BK79}:
\beq\begin{aligned}
B_{||} &\propto z^{-2} \textrm{ (magnetic flux conservation)},\\
B_{\perp} &\propto z^{-1} \textrm{ (magnetic energy conservation)}.
\end{aligned}\eeq
This causes the magnetic field to be predominantly toroidal in the radio-emitting parts of the jet (as evidenced by observations of radio and optical polarization of blazars, e.g., \citealt{2000Gabuzda, 2006Gabuzda, 2018Walker, 2022Issaoun, 2022Zhao}).

Internal shocks are driven along the jet axis into this changing configuration of magnetic fields. 
Close to the central engine, where $\vec{B}$ is randomly or poloidally oriented, this may result in quasi-parallel shocks (with $\vec{B}$ nearly parallel to the shock normal), which are known to be efficient particle accelerators in the standard Diffusive Shock Acceleration (DSA) mechanism \citep{SironiSpitkovsky2009, 2015Sironi}. Further away, the internal shocks likely become nearly exclusively perpendicular (with $\vec{B}$ perpendicular to the shock normal), the configuration that has long been considered to be inefficient for particle acceleration \citep{BegelmanKirk1990, 2013Sironi}. This prompted efforts in considering alternative particle energization mechanisms, such as magnetic reconnection \citep{Sironi2015}. It was recently found that perpendicular shocks may lead to particle acceleration when their surface is corrugated \identityone{\citep[][]{2022Demidem}}. Such a situation is not unlikely. Turbulence and density inhomogeneities within the jet can easily affect shock structure either directly or through secondary instabilities \citep[as has been shown for gamma-ray burst models, e.g.,][]{2016LopezCamara, 2020Gottlieb}. 

While corrugation offers a way to produce suprathermal particles in \identityone{quasi-perpendicular magnetized shocks} -- a natural configuration for internal shocks in jets -- its net effect on the downstream structure and radiative properties has not been investigated previously. Here, we assess whether the current phenomenological approach to compute spectral and timing properties of the internal shock model, which assumes one-dimentional collisions in their emission calculations \citep[e.g.,][]{Beloborodov2000, 2014Malzac, 2020Marino}, is still valid in the less trivial and arguably more realistic case of corrugated shocks.

Assuming that particles attain a power-law energy distribution in both cases, we find corrugated shocks to be up to $\sim10\%$ brighter than their flat counterparts. We acknowledge this to be a small correction in comparison to effects of other approximations made in internal shock models. Thus, we conclude that the existing internal shock models are entirely consistent with the behavior of 2D corrugated shocks, at least in the simplified MHD approach presented here. We caution that a detailed treatment of particle acceleration by corrugated shocks (either through test particles or fully PIC simulations) 
may well change this conclusion. 


Multiwavelength polarization studies are well known to reveal the structure of magnetic fields in jets \citep[e.g.][]{2000Gabuzda, 2006Gabuzda, 2022Issaoun}. With the recent launch of the \textit{Imaging X-ray Polarimetry Explorer} (\texttt{IXPE}, \citealt{Weisskopf2022}), studies have for the first time been able to reach the regions immediately behind the shock fronts. Our simulations allow us to make predictions regarding the level of polarization of jet synchrotron emission in the context of internal shock models. 

From the point of view of a single shell collision, our results allow us to apply the findings of \cite{2018Tavecchio}, leading to the prediction that a single two-shell collision in the region of toroidal magnetic field (far in the jet) should result in low optical and high X-ray polarization fraction. 
At the same time, a similar collision close to the central engine would occur at a much higher value of $B_{||}$ (as discussed above), resulting in highly polarized radiation in both spectral regions. 
However, the frequency of the peak of the synchrotron emission flux depends on the distance from the central engine ($\nu\appropto z^{-1}$, known as core-shift, \citealt{1998Lobanov, 2005Hirotani}), so regions close to the accretion disk will mainly radiate in X-ray, while those further out will be seen in optical/IR/radio bands. 
Thus, we can overall predict that the jet-integrated synchrotron emission from an internal-shock-powered jet should exhibit a highly-polarized X-ray emission and a weakly-polarized optical/IR emission. 
We are excited to see these expectations systematically investigated with multiwavelength polarimetric data.

\identityone{Our work expands the 1D picture of shell collisions, typically used with internal shock models, into 2D. A natural next step would be to investigate a fully 3D system. Instabilities may behave differently in 2D than in 3D. However, the only instability we observe in our runs is the Kelvin-Helmholtz instability, which, by its nature, remains locally two-dimensional. As reported by, e.g., \cite{2021Markwick} colliding flows may also exhibit radiative instabilities which do evolve differently in 3D. Still, due to the small radiative power of microquasar / AGN jets relative to their kinetic power, these instabilities should not have significant dynamical consequences on our systems of interest. Thus, we expect the downstream to behave similarly in 3D as it does in 2D. The main finding of our 2D work is that magnetic field is amplified in structured (as opposed to 1D) systems, leading to brighter synchrotron emission. It is difficult to say whether this amplification would be stronger or weaker in 3D configurations. However, a dynamo may operate in 3D in addition to the process we observe here, leading to further amplification of magnetic fields. Thus, we expect the increase in synchrotron brightness to be comparable or even larger in 3D than it is in 2D. Further work in 3D would be needed to address whether the conjectures we list above are correct.}


\begin{acknowledgments}
This research was performed with computing resources (Beskow, Dardel) generously provided by PDC Center for High Performance Computing at the KTH Royal Instutute of Technology, allocated through the Swedish National Infrastructure for Computing (SNIC). Nordita is supported in part by NordForsk. 
C.D. acknowledges support from NSF grant AST 1903335 and NASA grant NNX17AK55G.
A.V. acknowledges support from the Academy of Finland grants 309308 and 322779.
\end{acknowledgments}

\bibliographystyle{aasjournal}
\bibliography{references}

\appendix

\section{Parameter range estimation for \gx\ environment}\label{sect:gx_pars_derivation}

To obtain plasma parameters close to what is expected in \gx, we use existing theoretical models of the binary from the literature to provide lower and upper constraints on the values of interest, and then select values within the resulting ranges.

\paragraph{Lower limits:}
  
  \cite{2009Maitra} calculates the particle density of leptons at the jet ejection zone of \gx\ to be:
  \beq n_{\rm e}(r_{\rm ej}) \sim 4\times [10^{13.4},10^{15}]\,\textrm{cm}^{-3} \gtrsim 10^{14}\,\textrm{cm}^{-3}, \eeq
  where the factor of $4$ comes from transforming to post-shock density. The location of the ejection region is constrained within (their Table~2):
  \beq z_{\rm ej} \sim 33-114r_{\rm g}. \eeq
  Here, we will assume the jet to be roughly conical between the ejection zone and the IR zone of \cite{2011_Shidatsu} at $\sim10^4r_{\rm g}$. As the jet is likely convex while accelerating, this will likely underestimate the local particle number density:
  \beq n_{\rm e}(z) \propto z^{-2}, \eeq
  \beq n_{\rm e}(z_{\rm IR}) \gtrsim 1.3\times 10^{10}\,\textrm{cm}^{-3}. \eeq
  This gives us the local mass density:
  \beq \rho(z_{\rm IR}) \gtrsim n_{\rm e} m_{\rm p} \sim 4.2\times 10^{-14} \textrm{ g~cm}^{-3} \eeq
  and magnetization:
  \beq \sigma_{\rm j}(z_{\rm IR}) \simeq \frac{B^2/8\pi}{\rho(z_{\rm IR}) c^2} \lesssim 2.6. \eeq
  We can also calculate the plasma skin depth:
  \beq l_{\rm s} = \frac{c}{\omega_{pe}} = c\sqrt{\frac{m_e}{4\pi n_{\rm e} e^2}} \lesssim 3.4 \textrm{ cm}. \eeq

\paragraph{Upper limits:}
  
  As argued by \cite{2009Maitra}, the total jet power $P_{\rm j}$ is larger than bulk kinetic energy advection in protons by a factor of $\sim10$, as it also contains the energy later used to reheat the particle distribution (by, e.g., internal shocks) (see discussion in their Sect.~2.1 regarding the ``input power parameter'' $N_{\rm j}$ and its relation to total jet power of \citealt{2005_Markoff}). Interestingly, their $N_{\rm j}$ (see their Table 2) is actually very close in value to $P_{\rm j}$ from \cite{2018_Malzac} (however, as these are physically different models, this observation need not be necessarily meaningful). Given these findings, as well as results regarding jet power from \cite{2005Gallo} and \cite{2020Bright}, lowering the jet power $P_{\rm j}$ of \cite{2018_Malzac} by a factor of 10 can be used to estimate the local plasma parameters. Thus, in this paragraph, we will take $P_{\rm j}\sim 10^{37}\textrm{ erg} \textrm{s}^{-1}$.
  
  Assuming that jet power is fully dominated by the kinetic energy of protons (i.e., overestimating jet power):
  \beq P_{\rm j} \lesssim (\Gamma_{\rm j}-1) \rho c^2 \times v_{\rm j}  \times \pi(\theta_{\rm j} z)^2, \eeq
  \beq \rho(z) \lesssim \frac{P_{\rm j}}{(\Gamma_{\rm j}-1) v_{\rm j} c^2 \times \pi(\theta_{\rm j} z)^2}, \eeq
  \beq \sigma_{\rm j}(z) \gtrsim \frac{B^2 (\Gamma_{\rm j}-1) v_{\rm j} \times \pi(\theta_{\rm j} z)^2}{8\pi P_{\rm j}}. \eeq
  This, for the location of IR core from \cite{2011_Shidatsu}, gives us:
  \beq \rho(z_{\rm IR}) \lesssim 4.3\times 10^{-13} \textrm{ g~cm}^{-3}, \eeq
  \beq \sigma_{\rm j}(z_{\rm IR}) \gtrsim 0.25. \eeq
  Note that this is given in the frame of reference comoving with the jet at its bulk Lorentz factor, $\Gamma_{\rm j}\sim2$.
  
  Assuming charge neutrality, we can obtain electron number density in this frame from the mass density above:
  \beq n_{\rm e} \simeq \rho(z)/m_{\rm p} \gtrsim 2.6\times 10^{11}\,\textrm{cm}^{-3}. \eeq
  
  This gives a plasma skin depth of:
  \beq l_{\rm s} = \frac{c}{\omega_{pe}} = c\sqrt{\frac{m_e}{4\pi n_{\rm e} e^2}} \lesssim 1.0\textrm{ cm}. \eeq
  
\paragraph{Combined constraints}

Based on the considerations above, we can constrain the fluid parameters within \gx\ jet to be within the following ranges:
\beq \rho \in [4.2\times 10^{-14}, 4.3\times 10^{-13}]\textrm{ g~cm}^{-3}, \eeq
  \beq n_{\rm e} \in [1.3\times10^{10}, 2.6\times 10^{11}]\textrm{ cm}^{-3}, \eeq
  \beq \beta \sim 1, \eeq
  \beq \sigma \in [2.6, 0.25], \eeq
  \beq l_{\rm s} \in [3.4, 1.0] \textrm{ cm}, \eeq
  where $\rho$ denotes density, $n_{\rm e}$ -- electron density, $\beta$ -- plasma beta (defined as the ratio of particle to magnetic field energy density), $\sigma$ -- magnetization, and $l_{\rm s}$ is the plasma skin depth. Note that we have $\sim10^8-10^9$ of local skin depths per jet radius at our region of interest at $z_{\rm IR}\sim 10^4r_{\rm g}$.
  



\end{document}